\documentclass[a4paper]{sab}  

\usepackage{graphicx} 
\usepackage{txfonts,natbib} 
\bibpunct{(}{)}{;}{a}{}{,} 

\usepackage[T1]{fontenc} 
\usepackage{t1enc} 
\usepackage{aeguill} 
\renewcommand\ackname{Acknowledgements.} 
\renewcommand\keywordname{Keywords.} 
\renewcommand\figureptname{Figure}
\renewcommand\tableptname{Table}

\renewcommand\noteaddname{Note added to proofs.}
\idline{Boletim da Sociedade Astron\^omica Brasileira, {\bf 29}, XX-XX}{1}
\begin{document} 
\title{Systems Enginnering applied to spectroscopy of the ELTs} 
\subtitle{The Conceptual Design phase of GMACS}

\author{D. M. Faes\inst{1} \and A. Souza\inst{2,3} \and D. R. Bortoletto\inst{3} \and M. C. P. Almeida\inst{4}} 
\institute{
Instituto de Astronomia, Geof\'isica e Ci\^encias Atmosf\'ericas (IAG/USP) \\\email{daniel.gmt@iag.usp.br}  
\and
Universidade Federal de Minas Gerais (UFMG) 
\and 
Instituto Mau\'a de Tecnologia (IMT)
\and 
Universidade do Vale do Para\'iba (UniVap)
}

\date{Received XYZ; accepted XYZ} 

\Abstract {An important tool for the development of the next generation of extremely large telescopes (ELTs) is the Systems Engineering (SE). GMACS is the first-generation multi-object spectrograph working at visible wavelengths for the Giant Magellan Telescope (GMT). The aim is to discuss the application of SE in ground-based astronomy for multi-object spectrographs. For this, it is presented the SE of the GMACS spectrograph, currently on its Conceptual Design phase. SE provide means to assist the management of complex projects, and in the case of GMACS, to ensure its success when in operation, maximizing the scientific potential of GMT.}
{Uma ferramenta importante para o desenvolvimento da próxima geração de telescópios extremamente grandes (ELTs) é a Engenharia de Sistemas (SE). O GMACS é o espectrógrafo multi-objeto de primeira geração trabalhando em comprimentos de onda visíveis para o Telescópio Gigante de Magalhães (GMT). O objetivo é discutir a aplicação de SE em astronomia de solo para espectrógrafos de objetos múltiplos. Para isso, é apresentado o SE do espectrografo GMACS, atualmente em sua fase de Design Conceitual. SE oferece meios para auxiliar o gerenciamento de projetos complexos e no caso do GMACS, para garantir seu sucesso quando em operação, maximizando o potencial científico da GMT.}


\maketitle 

\section{Introduction}
The development of innovative scientific instrumentation has a number of challenges, involving its design, construction and long-term operation. Astronomy is no exception. An important tool for the development of the next generation of extremely large telescopes (ELTs) is the Systems Engineering (SE).

Multi-Object Spectroscopy (MOS) is one of the most demanding observational techniques in astronomy. The ELTs provide unique windows for scientific discoveries using MOS techniques. Good summaries of the science cases for MOS using the ELTs can be found in \citet{2006IAUS..232..204C}, \citet{2006SPIE.6272E..1XN}, and \citet{2015arXiv150104726E}.

In this context, it was proposed the \textit{Giant Magellan Telescope Multi-object Astronomical and Cosmological Spectrograph} (GMACS). GMACS is a multi-object spectrograph working at visible wavelengths for the GMT. See \citet{2014SPIE.9147E..20D} for a project status overview.

In Section~\ref{sec:SE} we present what is Systems Engineering and its importance for projects such as the ELTs. Section~\ref{sec:spec} containts a brief discussion of the challenges involved in the development of instrumentation for the ELTs, with emphasis in spectrographs. The Systems Engineering processes of the GMACS Conceptual Design phase are described in Section~\ref{sec:GMACS}. Our final remarks are in Section~\ref{sec:final}.

\section{Systems Engineering \label{sec:SE}}
Systems Engineering (SE) proposes a series of methodologies and practices to ensure the successful development and operation of systems. Historically, many of the SE processes application were in the aerospace industry and the defence industry \citep{incose}. However, nowadays SE has a broader scope of applications (e.g., Product-SE, Enterprise-SE, Service-SE, etc). For a discussion of the impact of SE in ground-based observatories, see \citet{2003SPIE.4837..166S}.

Some of the reasons that led to the implementation of SE methodology in complex projects are: (i) Limited product effectiveness; (ii) Results often unrelated to the actual needs; (iii) Serious delays in schedules; (iv) Excessive costs; (v) Bad development directions; (vi) Need for unification or standardization of practices created in different fields.

The early implementation of SE practices aims to guarantee a good understand of the needs and requirements of the system from concept to disposal. SE design methodology will widely consider the system life cycle, the needs of the final users and mitigate risks as early as possible by working closely with specialized engineers. 

\begin{figure}[!htbp]
  \centering
  \includegraphics[width=1.\linewidth]{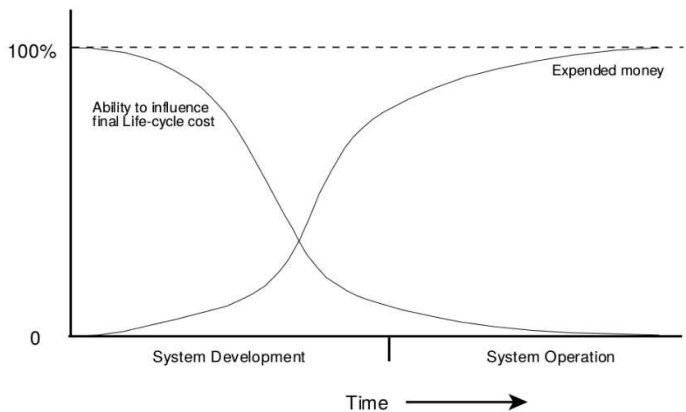}
  \caption{Declining ability to influence final system cost throughout a project (credit: \citealt{2003SPIE.4837..166S}).}
  \label{fig:0}%
\end{figure}

Figure~\ref{fig:0} shows the effectiveness of the application of SE throughout a project. This pattern has been observed in different projects from different domains. 

\section{Spectroscopy of the ELTs \label{sec:spec}}
The increasing cost of ground-based astronomy with the size of the telescope apertures, coupled with increased technical complexity are important reasons for observatories to seek the support of SE practices. Table~\ref{tab:1} contains data from the survey of \citet{van_belle_scaling_2004}, as well as GMT as a representative case of the cost of an ELT.

\begin{table}
\centering
\caption{Comparison of telescope projects. Costs are approximated and based on the buying power of the US dollar  in 2000.}
\begin{tabular}{ ccccc }
Telescope & Diameter & Start & First light & Cost \\ \hline
Magellan & (2x) 6.5m & 1994 & 2002 & \$135.4M \\
Keck II & 10m & 1991 & 1996 & \$85.9M \\
Keck I & 10m & 1985 & 1993 & \$139.1M \\
MMT & 6.5m & 1979 & 2000 & \$49.4M \\
GMT & 28m & 2007 & 2024 & \$1.000M \\
\end{tabular}
\label{tab:1}
\end{table}

In a very brief way, the main goal of the ELTs it to take spectra of targets that otherwise are only visible through images, like primordial (high-redshift) galaxies. Also, ELTs are the ideal tools to obtain high cadence of observations in transient events, such as transit of exoplanets. However, the construction of instrumentation for these and other MOS goals has a number of challenges. Here we list the needs of ELT spectrographs that should be addressed in a systemic way.

\subsubsection*{Scale up to keep FoV}
One of the main difficulties in the construction of spectrographs for ELTs is the physical size of the optics. By construction, the working f-number of reflector telescopes do not change considerably with its size. This means that the physical size of the generated images grows linearly with the diameter of the telescope. Table~\ref{tab:0} contains typical values for telescopes with f-number $\approx16$. For ELTs covering a reasonable wide field of view, one can expect images that are more than one meter in size!

\begin{table}
\centering
\caption{Comparison of the plate scale and a 10 arcmin image image of different telescopes aperture sizes and same effective f-number\,=\,16.5 (approx$.$ values).}
\begin{tabular}{ccc}
Diameter & Plate scale & Size of 10' \\
 (m) & (''/mm) & (mm) \\\hline
  3 & 4.0 & 150 \\
  10 & 1.3 & 450 \\
  30 & 0.4 & 1350 \\
\end{tabular}
\label{tab:0}
\end{table}

\subsubsection*{Competitive resolution and spectral coverage}
When the resolution of the generated spectra is considered, there is a similar impact. The main factor controlling the spectral resolution in terms of the size of the optics is the ratio between the diameters of the collimator and the telescope \citep{2007MNRAS.376.1099A}. Because it is very difficult to create large lenses, in first order the resolution of a given spectrograph is inversely proportional to the diameter of the telescope. 

\subsubsection*{High mechanical stability}
The size of the optics generate large instruments. Spectrographs in Cassegrain focus will need real-time mechanical actuators to correct mechanical flexure with gravity vector changes. This is true for GMACS, which will stand at the bottom of the telescope mount, and also to spectrographs in Nasmyth focus that need to rotate accordingly to the observed field. The total mass of the instruments increase the chances of inaccurate flexure corrections that can greatly degrade the efficiency and quality of the generated spectra.

\subsubsection*{Integration with AO capabilities}
The integration with adaptive optics resources simultaneously serves to identify and observe weak targets as well as it is an effective mechanism to increase the resolution of the generated spectra. The area of the primary ELT mirrors generate additional deflections for adaptive optics corrections, especially if it is considered multiple targets or a large field of view.

\subsubsection*{High throughput}
It seems to be a simple requirement, but high transparency is a challenge in a large system (which often uses internal mirrors to reduce its volume) and is still integrated into an adaptive optics system.


\section{GMACS as a subsystem of the GMT \label{sec:GMACS}}
As mentioned in Section~\ref{sec:SE}, SE methodology aims to address any issues of the project as early as possible. We describe here this methodology in more details, focusing in describing its tailored version as applied to GMACS.

The GMT System Engineering Framework defines the project hierarchy, overall scope of each project phase and highlights the common artifacts recommended to be used when implementing requirements flow-down, interface definition, risk analysis, planning, decision analysis and cost estimates \citep{2012SPIE.8449E..06M}.

GMT recommends this approach to all instrumentation groups. Similarly to GMACS, a novel systems engineering approach is being applied to the \textit{GMT-CfA Large Earth Finder} (G-CLEF) \citet{2014SPIE.9147E..8WP}.

\subsubsection*{Top-down approach}
The Top-down approach covers managerial and design practices. It is a way of managing and designing the project so that engineers can address first architectural aspects of the project without focus in detail. As more information becomes available, details will be addressed in the design. To start this SE seeks to capture all subsystems necessary, for that a PBS (Product Breakdown Structured) is developed together with engineers. The PBS will help manage the group, plan activities, organize the flow-down of requirements from system to subsystem.

\subsubsection*{Traceability of requirements and requirements flow-down}
The requirements flow-down at GMACS is responsibility of the system engineer with the support of specialized engineers and astronomers. It starts from the identification of scientific cases, operational aspects and constrains imposed by the observatory. From these, the first flow-down are written and the initial requirements that will guide the technical team captured. The Traceability from all identified aspects and the derived requirements are managed by the system engineer, using specific tools. The traceability of the requirements allows the system engineer, to be prepare to estimate the impact of changes, to know better the scope of the project, to justify decisions, besides estimate cost and schedule.

\subsubsection*{Record of decisions, knowledge management}
From the initial flown down of requirements many concepts are possible and the experience of the group and research are important to make decisions that will reduce the possibilities and narrow down those to the most likely options. The system engineers at GMACS oversees those decisions and participates to make sure the complete life-cycle is considered, instead of only performance and cost. All decisions are documented as Thread-off or technical notes from templates developed by SE team.

\subsubsection*{Cost and schedule estimates}
System engineering applied to cost and schedules estimates will consider the entire systems life cycle (development, fabrication, integration, validation, commissioning, operation, upgrade and disposal), in addition to social and environmental aspects that may influence in some stage of the system life-cycle.

Making long term estimates, is matter addressed statistically by SE tools. In the GMACS case, to deal with the possibilities and uncertainty of the six remaining phases until first light, the risk analysis will be used as input for the estimation tools.

\subsubsection*{Risk management}
\begin{figure}[!htbp]
  \centering
  \includegraphics[width=.8\linewidth]{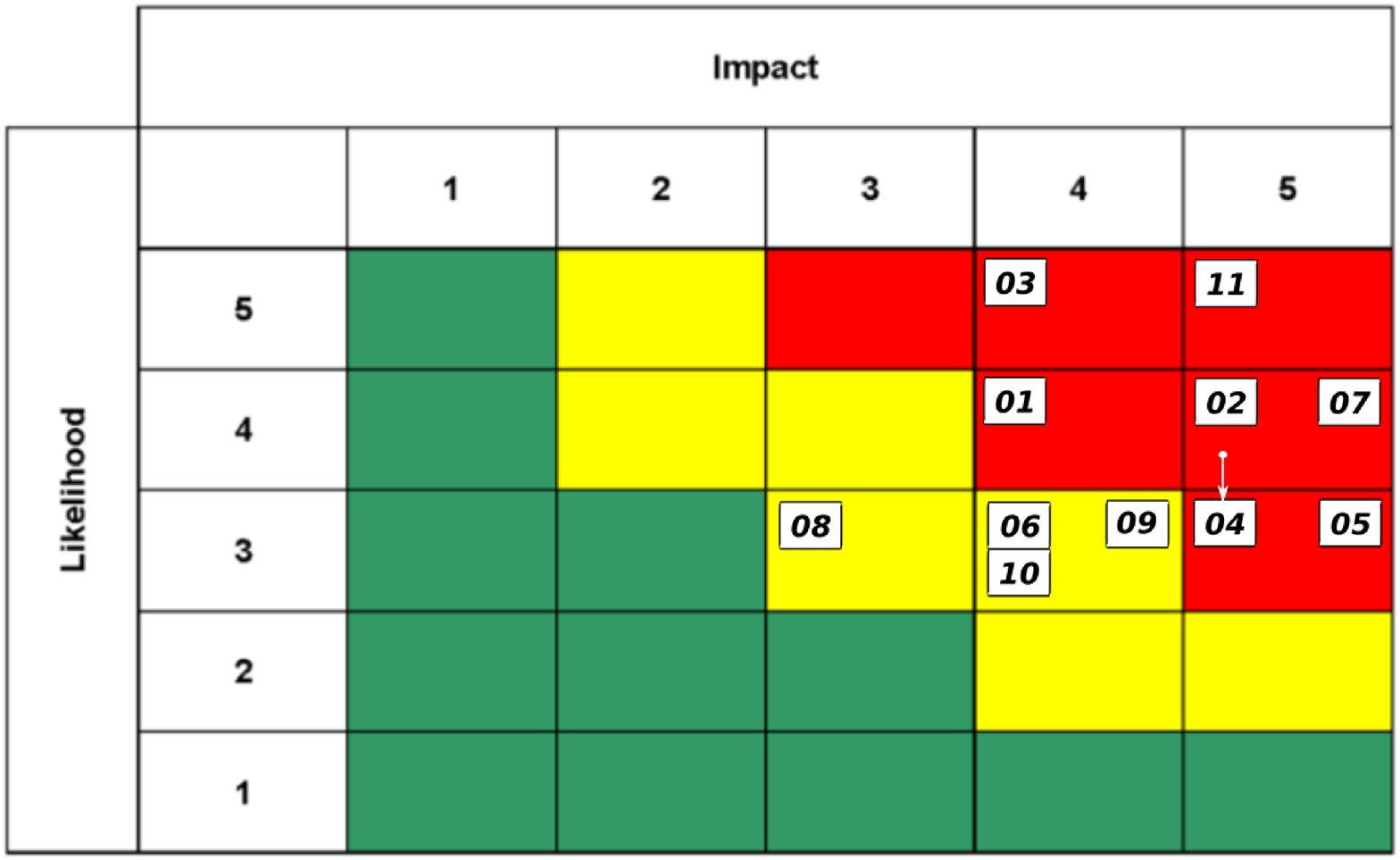}
  \caption{Example of risk matrix to identify and visualize the evolution of the risks. The higher the likelihood, more probable is its occurrence; the higher the impact, the greater the (negative) consequence in the project.}
  \label{fig:1}%
\end{figure}

Risk Management of the project allows the SE to identify technical and strategic risks and plan mitigation that can be applied in early stages. When applied at conceptual design, such as GMACS, the awareness of the risks allows to mitigate most of them during the tread-off and decision process. For GMACS, the expectation at the end of the conceptual design is to have all risks from the red area (Figure~\ref{fig:1}), moved to yellow and green, meaning that the risk will be much more manageable.

GMACS uses the same approach for risk management as GMT, only adapted for scaled reasons, at metrics for cost and schedule impacts and likelihood. Following that approach, all risks are classified as technical, cost, schedule and have the impacted requirement traced to it.

\subsubsection*{Interfaces}

Interface is one of the most challenging aspects that SE deals with. It requires communication, organization, discipline and knowledge of the overall aspects of the system, its subsystems, operation and environment. For a conceptual design like GMACS, top-down approach and bottom-up approach need to be combined to consider all interfaces. Top-down allows the identification of interfaces from a wide point of view, considering observatory aspects, such as operation, facility instruments and AIT. Bottom-up complements by allowing the identification of interfaces that depends of subsystems solutions. In order to coordinate both approaches, good practices of requirements traceability and knowledge management need to be followed, which includes good communication between all stakeholders that SE needs to be prepared to facilitate.

\section{Final Remarks \label{sec:final}}
This work addresses the objectives of SE in complex systems and how SE is proposed in the GMACS project. This is contextualized within SE processes for GMT, and the focus is on the challenges of the multi-object spectroscopy technique for the ELTs need to overcome. From a broader perspective, it is pointed out how SE methods can assist the development of complex projects and maximize the scientific potential of big experiments, such as the ELTs.

\begin{acknowledgements}
DMF acknowledges support from FAPESP grants 2011/51680-6 and 2016/16844-1.
\end{acknowledgements}

\bibliographystyle{apalike}
\bibliography{bib}

\end{document}